\documentclass[12pt]{amsart}

\newcommand {\fgl}    {{\mathfrak{gl}}}  %
\newcommand {\fosp}   {{\mathfrak{osp}}}
\newcommand {\fsl}    {{\mathfrak{sl}}}
\newcommand {\fq}     {{\mathfrak{q}}}     

\newcommand {\Cee}    {{\mathbb  C}}
\newcommand {\Zee}    {{\mathbb  Z}}

\def \opname#1#2%
  {\expandafter\newcommand \csname #1\endcsname {{\mathop{#2}\nolimits}}}

\newcommand{\rmname}[1]
  {\expandafter\newcommand \csname #1\endcsname {{\operatorname{#1}}}}

\rmname{Hom}
\rmname{ind}

\rmname{Ker}
\rmname{Coker}

\newcommand {\eps} {\varepsilon}

\newcommand {\tto} {\longrightarrow}

\newcommand {\secno} {}

\newtheorem{Theorem}{\secno Theorem}

\newenvironment {th*}[1]
    {\gdef\thname{#1} \begin{thn}}%
    {\end{thn}}
\newtheorem{thn}[Theorem] {\thname}

\theoremstyle{definition}

\newenvironment {ex*}[1]
    {\gdef\thname{#1} \begin{exn}}%
    {\end{exn}}
\newtheorem{exn}[Theorem]{\thname}

\theoremstyle{remark}

\newenvironment {rem*}[1]
    {\gdef\thname{#1} \begin{remn}}%
    {\end{remn}}
\newtheorem{remn}[Theorem]{\thname}

\newcommand {\ssec}{\subsection*}

\begin{document}
	
\title[The Index Theorem on Supermanifolds]{The Index Theorem\\ for 
Homogeneous Differential Operators\\ on Supermanifolds}

\author{Dimitry Leites}

\address{Department of Mathematics, University
of Stockholm, Roslagsv.  101, Kr\"aftriket hus 6, SE-106 91, Stockholm,
Sweden; mleites@mathematik.su.se}

\thanks{Financial support of IHES in 1986; MPI, Bonn, in 1987 and
Swedish NFR afterwards is gratefully acknowledged.  I am thankful to
M.~Shubin and A.~Dynin (one of the pioneers of the index theory) for
lending a hand during my first attemts to write the index theorem on
supermanifolds; to J.~Bernstein and I.~Penkov who taught me a lot I
needed in this work; to I.~M.~Gelfand and B.~Sternin for
encouragement. }



\begin{abstract} In mid 60s Bott proved that (1) for homogeneous,
i.e., $G$-invariant, elliptic differential operators acting in the
spaces of sections of induced representations of $G$ over $G/H$ the
index theorem reduces to the Weyl character formula and (2) the index
of an equivariant elliptic operator {\it does not depend on the
operator} but on the representations.  Here the same theorem is
formulated for the unitary supergroup $G=U(p|q)$.  For atypical
representations the character formula does not reduce to that for the
Lie group underlying the supergroup $G$ and this contradicts a
statement of Rempel and Schmitt on index on supermanifolds
(Pseudodifferential operators and the index theorem on supermanifolds. 
Seminar Analysis, 1981/82, 92--131, Akad.  Wiss.  DDR, Berlin, 1982;
id., Pseudodifferential operators and the index theorem on
supermanifolds.  Math.  Nachr.  111 (1983), 153--175).
\end{abstract}

\maketitle

\ssec{Prehistory} Israel Gelfand conjectured that since the index of
any elliptic operator does not vary under deformations, it should be
possible to express it in terms of the topological data of the
problem.  Atiyah and Singer were the first to prove this conjecture. 
Since then, there were given several proofs of this conjecture until
finally Quillen not only noticed a manifestly \lq\lq superish" nature
of the index theorem --- the index is the supertrace of an operator
--- but in two papers that drew a lot of attention to the then new
field, supermanifold theory, gave an outline of the proof whose
clarity was due to this nature.  Getzler did a lot on elaboration and
elucidation of the mechanism of this proof see \cite{Q} and
\cite{BGV}.  (Of particular interest are papers by Alvarez-Gom\'e and
Witten, see \cite{GSW}, who link this, super, approach to physics.)

Quillen and his first followers, even Witten, only used the most
superficial features of super-manifold theory: it was sufficient for
the classical index theorem they were interested in.  For the
researchers of supermanifold theory proper it is tempting to attack a
more general problem: to translate the index theorem and related
topics, like the Riemann--Roch--Hirzebruch--Grothendieck theorem, from
manifolds to supermanifolds.  How to perform such a generalization is,
however, quite unclear even after Manin's contribution towards
solution \cite{M}.

One of the obstacles: the de Rham cohomology of a (smooth)
supermanifold coinsides with that of the underying manifold.  The
inadequacy of de Rham cohomology as an instrument for description of
supermanifolds is manifest: the dimension of a vector superspace or a
supermanifold is an element of the ring $\Zee [\eps]/(\eps^2-1)=\Zee
\oplus\Zee \eps$, so attempts were made to label cohomology, as well
as homology, by a {\it pair} of integers, rather than by a single one.

The two most successful approaches are those expressed in: 

(1) papers by Manin and his students on {\it homology} of
supergrassmannians: the idea to label Schubert supercells with
elements of an analogue of the Weyl group.  Several versions of the
exposition was written; for a review see the translation \cite{M} of
VINITI's collection.  This review contains also an alternative and in
some cases different description of super analog of the Weyl group,
due to Penkov and Skornyakov.  This rival approach, later improved yet
further \cite{PS}, is especially good for superization of the
Borel--Weil--Bott--Kostant-Bernstein--Gelfand--Gelfand theorem.

This approach, however, only embraced cell decomposition of quotients 
of simple Lie supergroups modulo parabolic subsupergroups, not the 
general supermanifolds.  And nobody had yet succeded to introduce {\it 
cohomology} of supermanifolds labelled with two numbers except 
Shander:

(2) Shander's integration theory concerning complexes of 
pseudoforms, see \cite{L4}, no.  33, Chapter 5.

Another obstacle for superization: one of the most conventional
formulations of the index theorem --- in terms of fixed points ---
baffles anyone who tries to formulate it on a supermanifold of purely
odd dimension or for an odd differential operator.  Again Shander is
the only one who managed to see some ways out of this predictment, see
\cite{SH}.

Summing up, one was tempted to conjecture that the index theorem on 
supermanifolds reduces to that on underlying manifolds.  Indeed, such 
a theorem was soon published, see \cite{RS}.  

I never believed in this theorem.  The reason for my incredulity was
based on two facts:

(a) still another formulation of a particular instance of the index 
theorem, due to Bott \cite{B}, in the case of an {\it homogeneous}, 
i.e., $G$-invariant, differential operator on the coset space $G/H$ 
states that the index theorem reduces to the Weyl character formula 
for the irreducible representations of the simple Lie group $G$;

(b) the superized Weyl character formula reduces to the representation
of the underlying Lie algebra for typical representations only,
otherwise it does not and the character of atypical representations is
given by the so-called Bernstein--Leites character formula for
$\fsl(1|n)$ \cite{BL} and $\fosp(2|2n)$ \cite{L3}, respectively; for
the most general form of the character formula see \cite{PS},
\cite{S}.  (Recall also earlier descriptions of the typical
representation \cite{Be} for $\fgl$ and $\fosp$; \cite{K} for all
simple finite dimensional Lie superalgebras with Cartan matrix;
\cite{P} for the series $\fq(n)$ and the review \cite{L3} for all
representations, not only typical ones of the other series).  The
atypical representations are the most interesting ones: the identity
one, the adjoint one, their symmetric powers, etc.

In \cite{L1} I suggested to superize the index theorem along the lines
of \cite{B} but for various reasons the project was put aside.  Recent
discussions with B.~Sternin and I.~M.~Gelfand encouraged me to reread
\cite{B} to see what can be sulvaged under superization; this note is
a straightforward answer; conjecturally there exists an algebraic
proof fitting any character formula.
 
I will confine myself to the case of unitary supergroups only.  This,
I am sure, is a purely technical restriction: the idea of the proof of
the main theorem here follows word-for-word that of \cite{B}, where
the compactness of $G$ is crucial.  For the Lie groups this does not
restrict the generality, since every simple Lie group has a compact
form.  Contrarywise, almost no simple complex Lie supergroup has a
compact form and there are many compact superspaces $G/H$ with a
noncompact $G$, see \cite{Se}.  Nevertheless, I am sure that the
compactness of $G$ is beside the point here.  The statement, being of
purely algebraic nature, should be proved accordingly.

\ssec{Digest of Bott's result} Let $E$ and $F$ be smooth complex
vector bundles over a smooth compact manifold $M$; let $D:\Gamma (E)
\tto \Gamma (F)$ be an {\it elliptic} operator, see \cite{BGV}.  Set
$$
\ind (D)=\dim\Ker (D)-\dim\Coker(D). 
$$
It is now well-known, that if $D$ is elliptic, then $\ind (D)$ is 
finite (\cite{BGV}).

Let $M=G/H$, where $G$ is a compact connected Lie group (in the proof
we also need $\pi_{1}(G)$ to be without 2-torsion), $H$ a connected
Lie subgroup of the same rank as $G$.  Let $E$ and $F$ be induced by
representations of $H$.  We will say that $D$ is {\it homogeneous} if
it is $G$-invariant.  Passing from manifolds to supermanifolds we
require the above for their respective underlying manifolds.

Define the ring $R(G)$ of virtual $G$-modules as follows.  Let $R(G)$,
as additive group, be generated over ${\Zee}$ by the full set
$\{M_{\lambda}\}_{\lambda\in\Lambda}$ of irreducible $G$-modules.  The
representative of $M_{\lambda}$ in $R(G)$ is called its {\it class}
and denoted by $[M_{\lambda}]$.  The ring structure of $R(G)$ comes
from tensoring: $[A][B]=[A\otimes B]$.  The map $[\cdot]: M\to [M]$
should possess the following properties of which only 4) is new as
compared with \cite{B}:

1) any additive function $f: \text{G-mod}\tto \text{Ab}$ on the
category of $G$-modules with values in the category of abelian groups
$\text{Ab}$ can be extended to a function $f: R(G)\tto \text{Ab}$ such
that $f(M)=f([M])$ and if the sequence $0\tto A\tto B\tto C\tto 0$ is
exact, then $f([B])=f([A])-f([C])$;

2) any group homomorphism $\varphi: H\tto G$ extends to a ring 
homomorphism $\varphi^*: R(G)\tto R(H)$;

3) the map $M\to\dim M$ (here $\dim$ is the superdimension for 
modules over supergruops) can be 
lifted to a ring homomorphism $\dim: R(G)\tto \Zee[\eps]$.

4) $R(G)$ is a $\Zee[\eps]$-module with the action of $\eps$ given by
the formula $\eps[M]=[\Pi(M)]$.

Set $\chi (D)=[\Ker D]-[\Coker D]$.  This is a refined index, since
$\ind (D)=\dim \chi (D)$.

Denote by $\hat R(G)$ a completion of $R(G)$, the ring of formal power
series on the same generators as $R(G)$.  Let us extend the
intertwining number of two $G$-modules
$$
(A, B)=\dim_{\Cee}\Hom_G(A, B)
$$
to a symmetric bilinear form $<\cdot, \cdot>_G$ on $R(G)$ by setting
$<M_{\lambda}, M_{\nu}>_G=\delta_{\lambda, \nu}$.

Given a group homomorphism $i: H\tto G$, define the formal inducing
$i_*: R(H)\tto \hat R(G)$ by seting
$$
i_*:[A]\mapsto\sum\limits_{\lambda}<i^*(M_{\lambda},
[A]>_H[_{\lambda}], \quad \text{ for any }A\in Ob~\text{H-mod},
M_{\lambda}\in Ob~\text{G-mod}.
$$

For any homogeneous elliptic operator $D:\Gamma (E) \tto \Gamma (F)$,
where $E$ and $F$ are induced from $H$-modules $K$ and $L$,
respectively, define the homogeneous symbol of $D$ to be
$s(D)=[K]-[L]\in R(H)$.

\begin{Theorem} {\em 1)} $\chi (D)=i_*(s(D))$.

{\em 1)} For any $G$-module $M$ there exists a homogeneous elliptic
operator $D_{M}: \Gamma(E)\tto \Gamma(F)$ with $\chi(D_{M})=[M]$. 
\end{Theorem}

The beauty of this theorem is twofold: the right hand side of 1) is
the Weyl character formula of $M$ for $D=D_{M}$ and thanks to property
1) of the function $[\cdot]$ we see that $[\Ker D]-[\Coker D]=[E]-[F]$
does not depend on $D$!  (Indeed, the sequence $0\tto \Ker D\tto E\tto
F\tto \Coker D\tto 0$ is exact.)

To superize Bott's result, we only have to replace in the above
definitions groups with supergroups and dimension with superdimension,
so the target ring in property 3) is $\Zee[\eps]/(\eps^2-1)$.  To
follow Bott's proof, we need to integrate over a supermanifold with
compact base which is only possible for the unitary supergroup.  I am
sure, however, that there is another, algebraic, proof of this
theorem.


\begin{thebibliography}{9999}

\bibitem[Be]{Be} 
Berezin, F. Representations of the supergroup $U(p, q)$.  Funkcional.  
Anal.  i Prilozhen.  10 (1976), n.  3, 70--71 (in Russian); id., {\em
Analysis with anticommuting variables}, Kluwer, 1987
\bibitem[BL]{BL}
Bernstein J., Leites D., A formula for the characters of the 
irreducible finite-dimensional representations of Lie superalgebras of 
series $\fgl$ and $\fsl$.  (Russian) C. R. Acad.  Bulgare Sci.  
33, no.  8, 1980, 1049--1051
\bibitem[B]{B} 
Bott, R. The index theorem for homogeneous differential operators, In: 
{\it Differential and combinatorial topology} (A symposium in honor of 
Marston Morse), 1965, Princeton Univ.  Press, Princeton, NJ, 167--186
\bibitem[BGV]{BGV} 
Berligne R., Getzler E., Vergne M., {\em Heat kernels and Dirac 
operators}.  Grundlehren der Mathematischen Wissenschaften 
[Fundamental Principles of Mathematical Sciences], 298.  
Springer-Verlag, Berlin, 1992.  viii+369 pp.
\bibitem[GSW]{GSW} 
Green, M.; Schwarz, J.; Witten, E., {\em Superstring theory}.  Second 
edition.  Vol.  1.  Introduction.  Cambridge Monographs on 
Mathematical Physics.  Cambridge University Press, Cambridge-New York, 
1988.  x+470 pp; Vol.  2.  Loop amplitudes, anomalies and 
phenomenology.  ibid., xii+596 pp.  %(Russian translation from the 
%first edition: {\cyr Teoriya superstrun}.  ``Mir'', 
%Moscow, 1990. 
\bibitem[K]{K}
Kac V.G., Characters of typical representations of classical Lie 
superalgebras.  Commun.  Alg.  v. 5, 1977, 889--897
\bibitem[L1]{L1} 
Leites, D., {\em The supermanifold theory}, 1983, Karelia Branch of
the USSR Acad.  Sci., Petrozavodsk, USSR , 200 pp.  (in Russian; for
an expanded English version see [L4])
\bibitem[L2]{L2}
Leites, D., Quantization and supermanifolds.  (Supplement 3).  In: F.
Berezin, M. Shubin, {\em The Schr\"odinger equation}, 1991, Kluwer,
Dodreht, 483--519
\bibitem[L3]{L3}
Leites D., A formula for the characters of the irreducible
finite-dimensional representations of Lie superalgebras of series $C$. 
(Russian) C. R. Acad.  Bulgare Sci.  33, no.  8, 1980, 1053--1055;
id., Lie superalgebras.  In: Current problems in mathematics, Vol. 
25, Itogi Nauki i Tekhniki, Akad.  Nauk SSSR, Vsesoyuz.  Inst. 
Nauchn.  i Tekhn.  Inform., Moscow, 1984, 3--49 (in Russian; the
English translation: J. Soviet Math., v.  30 (6), 1985, 2481--2512)
\bibitem[L4]{L4} 
Leites, D. (ed.) {\em Seminar on supermanifolds}, preprinted as 
Reports of Department of Mathematics of Stockholm University, no. 
1-34, 1988-91, 2100 pp. 
\bibitem[M]{M}
Manin Yu.  (ed.)  Current problems in mathematics.  Newest results,
Vol.  32, Itogi Nauki i Tekhniki, Akad.  Nauk SSSR, Vsesoyuz.  Inst. 
Nauchn.  i Tekhn.  Inform., Moscow, 1988, 3--25.  The English
translation: J. Soviet Math.  vol.  51, 1990, no.  1, 2069--2083
\bibitem[P]{P}
Penkov, I. Characters of typical irreducible finite-dimensional 
$\fq(n)$-modules.  (Russian) Funktsional.  Anal.  i Prilozhen.  20 
(1986), no.  1, 37--45
\bibitem[PS]{PS}
Penkov I., Serganova V., Generic irreducible representations of finite 
dimensional Lie superalgebras.  Internat.  J. Math.  5, 1994, 389--419
\bibitem[Q]{Q}
Quillen D., Superconnections and the Chern character.  Topology 24 
(1985), no.  1, 89--95; Mathai V., Quillen D., Superconnections, Thom 
classes, and equivariant differential forms.  Topology 25 (1986), no.  
1, 85--110; Quillen D., Superconnection character forms and the Cayley 
transform.  Topology 27 (1988), no.  2, 211--238
\bibitem[RS]{RS}
Rempel, S.; Schmitt, T., Pseudodifferential operators and the index 
theorem on supermanifolds.  Seminar Analysis, 1981/82, 92--131, Akad.  
Wiss.  DDR, Berlin, 1982; id., Pseudodifferential operators and the 
index theorem on supermanifolds.  Math.  Nachr.  111 (1983), 153--175
\bibitem[Se]{Se}
Serganova V., Classification of simple
real Lie superalgebras and symmetric superspaces.  (Russian)
Funktsional.  Anal.  i Prilozhen.  17 (1983), no.  3, 46--54.  English
translation: Functional Anal.  Appl.  17 (1983), no.  3, 200--207
\bibitem[S]{S}
Serganova V., Characters of irreducible representations of simple Lie 
superalgebras.  Proceedings of the International Congress of 
Mathematicians, Vol.  II (Berlin, 1998).  Doc.  Math.  1998, Extra 
Vol.  II, 583--593
\bibitem[SH]{SH} 
Shander V., Orbits and invariants of the supergroup ${\rm GQ}\sb n$.  
(Russian) Funktsional.  Anal.  i Prilozhen.  26, no.  1, 1992, 69--71; 
translation in Functional Anal.  Appl.  26 no.  1, 1992, 55--56; a 
detailed exposition in xxx.lanl.gov math.RT/9810112

\end{thebibliography}
\end{document}